\documentclass[prl,a4paper,10pt,twocolumn,showpacs,floatfix,superscriptaddress,amsmath,amsfonts,amssymb,preprintnumbers,longbibliography,citeautoscript]{revtex4-1}
\usepackage[T1]{fontenc}
\usepackage[utf8]{inputenc} %utf8 or latin1
\usepackage{dcolumn,graphicx,color,booktabs,microtype,afterpage}
\graphicspath{{./}{figure/}}
\usepackage[charter,greekuppercase=italicized]{mathdesign}

\makeatletter
\renewcommand\section{\@startsection {section}{1}{\z@}%
   {-2.5ex \@plus -1ex \@minus -.2ex}%
   {0.4ex \@plus.2ex}%
  {\normalfont\large\bfseries}}
\makeatother
\makeatletter
\renewcommand\subsection{%
  \@startsection{subsubsection}{3}{\z@}{3.25ex \@plus1ex \@minus.2ex}%
  {-1em}{\normalfont\normalsize\bfseries}}
\makeatother

\renewcommand{\figurename}{Figure}
\renewcommand{\tablename}{Table}
\makeatletter\renewcommand{\fnum@figure}[1]{\textbf{\figurename~\thefigure~|\ }}\makeatother
\makeatletter\renewcommand{\fnum@table}[1]{\tablename~\thetable.}\makeatother

\newcount\hh \newcount\mm
\hh=\time \divide\hh by 60
\mm=\hh \multiply\mm by 60 \mm=-\mm
\advance\mm by \time
\def\now{\number\hh:\ifnum\mm<10{}0\fi\n++umber\mm}

\definecolor{dgreen}{RGB}{0,127,0}

\begin{document}

\makeatletter\renewcommand{\ps@plain}{%
\def\@evenhead{\hfill\itshape\rightmark}%
\def\@oddhead{\itshape\leftmark\hfill}%
\renewcommand{\@evenfoot}{\hfill\small{--~\thepage~--}\hfill}%
\renewcommand{\@oddfoot}{\hfill\small{--~\thepage~--}\hfill}%
}\makeatother\pagestyle{plain}

\title{~\vspace{-1ex}\\ High-$T_\mathrm{c}$ superconductivity in undoped ThFeAsN}

\author{T.\,Shiroka}\email[\vspace{8pt}]{tshiroka@phys.ethz.ch}
\affiliation{Laboratorium f\"ur Festk\"orperphysik, ETH H\"onggerberg, CH-8093 Z\"urich, Switzerland}
\affiliation{Paul Scherrer Institut, CH-5232 Villigen PSI, Switzerland}

\author{T. Shang}\thanks{tian.shang@psi.ch}
\affiliation{Laboratory for Scientific Developments and Novel Materials, Paul Scherrer Institut, CH-5232 Villigen PSI, Switzerland}
\affiliation{Swiss Light Source, Paul Scherrer Institut, CH-5232 Villigen PSI, Switzerland}
\affiliation{Institute of Condensed Matter Physics, École Polytechnique Fédérale de Lausanne (EPFL), CH-1015 Lausanne, Switzerland}

\author{C. Wang}
\affiliation{Department of Physics, Shandong University of Technology, Zibo 255049, China}

\author{G.-H. Cao}
\affiliation{Department of Physics and State Key Lab of Silicon Materials, Zhejiang University, Hangzhou 310027, China}
\affiliation{Collaborative Innovation Centre of Advanced Microstructures, Nanjing 210093, China}

\author{I. Eremin}
\affiliation{\textsuperscript{}Institut für Theoretische Physik III, Ruhr-Universität Bochum, D-44801 Bochum, Germany}
\affiliation{\textsuperscript{}Institute of Physics, Kazan Federal University, 420008 Kazan, Russian Federation}

\author{H.-R.\,Ott}
\affiliation{Laboratorium f\"ur Festk\"orperphysik, ETH H\"onggerberg, CH-8093 Z\"urich, Switzerland}
\affiliation{Paul Scherrer Institut, CH-5232 Villigen PSI, Switzerland}

\author{J.\,Mesot}
\affiliation{Laboratorium f\"ur Festk\"orperphysik, ETH H\"onggerberg, CH-8093 Z\"urich, Switzerland}
\affiliation{Paul Scherrer Institut, CH-5232 Villigen PSI, Switzerland}
\affiliation{Institute of Condensed Matter Physics, École Polytechnique Fédérale de Lausanne (EPFL), CH-1015 Lausanne, Switzerland}

\begin{abstract}
\noindent
Unlike the widely studied $Re$FeAsO series, the newly discovered iron-based superconductor 
ThFeAsN exhibits a remarkably high critical temperature of 30 K, without chemical doping or 
external pressure. Here we investigate in detail its magnetic and superconducting properties 
via muon-spin rotation/relaxation ($\mu$SR) and nuclear magnetic resonance (NMR) techniques 
and show that ThFeAsN exhibits strong magnetic fluctuations, suppressed below $\sim 35$\,K, 
but no magnetic order. This contrasts strongly with the $Re$FeAsO series, where stoichiometric 
parent materials order antiferromagnetically and superconductivity appears only upon doping.
The ThFeAsN case indicates that Fermi-surface modifications due to structural distortions and 
correlation effects are as important as doping in inducing superconductivity. The direct competition 
between antiferromagnetism and superconductivity, which in ThFeAsN (as in LiFeAs) 
occurs  at already zero doping, may indicate a significant deviation of the  $s$-wave 
superconducting gap in this compound from the standard $s^{\pm}$ scenario.
\end{abstract}

%%%% Alternative abstract %%%%
%To understand the remarkably high critical temperature of 30\,K of the ThFeAsN 
%superconductor which, unlike the widely studied $Re$FeAsO series, is achieved without 
%chemical doping or external pressure, we investigate its magnetic and superconducting 
%properties via muon-spin rotation/relaxation ($\mu$SR) and nuclear magnetic 
%resonance (NMR) techniques. In other $Re$FeAsO iron pnictides, the stoichiometric 
%parent material orders antiferromagnetically and superconductivity appears only upon 
%doping. In ThFeAsN, zero-field $\mu$SR and NMR spin-lattice relaxation suggest strong 
%magnetic fluctuations, suppressed below $T^{*} \sim 35$\,K, but no magnetic order. 
%At low temperatures, transverse-field $\mu$SR data indicate a superconducting state 
%compatible with a two-gap $s$-wave (or an anisotropic $s$-wave) scenario, in turn 
%consistent with NMR spin-lattice relaxation data. The competition between antiferromagnetism 
%and superconductivity, which in ThFeAsN (as in LiFeAs) occurs already at zero doping, 
%may indicate a significant deviation of the  $s$-wave superconducting gap in this 
%compound from the standard $s^{\pm}$ case.


\maketitle\enlargethispage{3pt}

In the vast class of iron-based superconductors (IBS), only very few are 
superconductors in their original stoichiometric composition of compensated metals.
Among them are LaFePO, with a critical temperature $T_\mathrm{c} \simeq 4$\,K \cite{Kamihara2006}, 
LiFeAs, with $T_\mathrm{c}=18$\,K \cite{Tapp2008}, and FeSe with $T_\mathrm{c}=8$\,K 
\cite{Hsu08}. Most of the other materials, including LaFeAsO, are antiferromagnets 
with $T_{\mathrm{N}}$ of the order of 100\,K \cite{Kamihara2008}. 
Given the itinerant character of charge carriers in IBS, superconductivity 
can be achieved via Fermi-surface tuning in two possible ways: through 
isovalent substitution of ions with different radii, or by injection of electrons 
or holes in the Fe planes of the magnetically-ordered parent compounds 
\cite{Wen2008,Sadovskii2008,Paglione2010,Johnston2010,Si2016}.
In particular, for many 1111 materials, including LaFeAsO, enhanced 
critical temperatures are only achieved by F- or H-doping 
\cite{Wen2008,Sadovskii2008,Paglione2010,Johnston2010,Si2016,Muraba2014,Muraba2015}. 
For most of the stoichiometric compensated-metal IBS the superconducting properties 
deviate significantly from those where superconductivity is induced by doping. For 
example, it has been claimed that the tendency towards antiferromagnetic 
(AF) order favors the so-called sign-changing (between electron- and hole pockets)
$s^{\pm}$-wave symmetry of the superconducting state, which originates from enhanced 
repulsive interactions between electron and hole bands \cite{Mazin2008,Chubukov2008}. 
This interaction generally favors antiferromagnetism, but once disorder, pressure, 
or doping suppress the long-range AF order, the $s^{\pm}$-wave superconductivity emerges. 
By contrast, for LaFePO, LiFeAs, FeSe and, as we show below also for ThFeAsN, 
where superconductivity occurs in stoichiometric compounds without imposing pressure 
or doping, there are strong indications that, due to orbital effects, the 
superconducting-gap symmetry deviates significantly from the
$s^{\pm}$ scenario \cite{Kuroki2009,Ahn2014,Sprau2016,Nourafkan2016}.

Very recently, the undoped compound ThFeAsN was found to exhibit 
an onset of superconductivity (SC) at a remarkably high $T_\mathrm{c}$ of 30\,K, as established 
by magnetic susceptibility and electrical resistivity measurements \cite{Wang2016}. 
Considering that oxygen and selenium are typical ingredients of 
stoichiometric IBS materials, the absence of chalcogen 
elements in ThFeAsN is remarkable. 

Since energy-dispersive X-ray (EDX) analyses of the synthesized 
polycrystalline TheFeAsN material indicate no distinct O-for-N substitutions, 
it is of obvious interest to establish why, despite the 
lack of formal or incidental doping, a fairly high $T_\mathrm{c}$ is achieved in this case 
and to what extent the magnetic correlations support, interfere, or compete with 
superconductivity.
Pure compounds such as ThFeAsN, for which reliable band-structure 
calculations are available, represent ideal candidates also for testing 
contending theories of iron-based superconductors.

In the following, we report on the detailed microscopic investigation of ThFeAsN 
by employing muon-spin rotation/relaxation ($\mu$SR) and nuclear magnetic 
resonance (NMR) techniques, which are used to probe the magnetic and 
electronic properties of ThFeAsN at a local level. 
As we show below, the experimental results indicate that in this material 
strong spin fluctuations precede the onset of the superconducting phase 
and compete with it. 
The appearance of superconductivity in ThFeAsN in its pristine form 
seems ultimately related to an appropriately tuned electronic band structure,
which also clarifies the rare peculiarity of this system.
\begin{figure}[t]
\includegraphics[width=0.85\columnwidth]{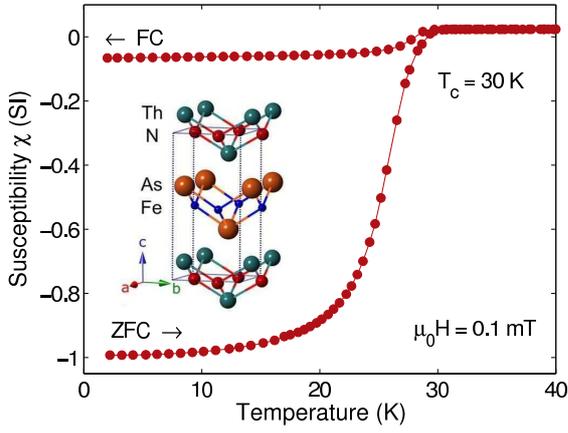} %{ThFeAsN_magn_struct}
\caption{\label{fig:magnetization}\textbf{Magnetic susceptibility of ThFeAsN.} 
Temperature dependence of the zero field-cooled (ZFC) and 
field-cooled (FC) dc susceptibility measured at  $\mu_0H = 0.1$\,mT. The sample shows a 
sizable diamagnetic response and $T_\mathrm{c} = 30$\,K. Inset: structure of ThFeAsN showing the 
ThN and FeAs planes (adapted from \cite{Wang2016}).
}
\end{figure}

\section{Results}
\subsection{Probing the intrinsic magnetism via zero-field $\mu$SR.} 
Preliminary susceptibility measurements $\chi(T)$ below 40\,K were 
used to detect the onset of bulk superconductivity. As shown in 
Fig.~\ref{fig:magnetization}, both ZFC and FC data indicate a 
SC transition at $T_\mathrm{c} = 30\pm0.5$\,K, in good agreement with the 
originally-reported value \cite{Wang2016}.

\begin{figure}[t]
\centering
\includegraphics[width=0.85\columnwidth]{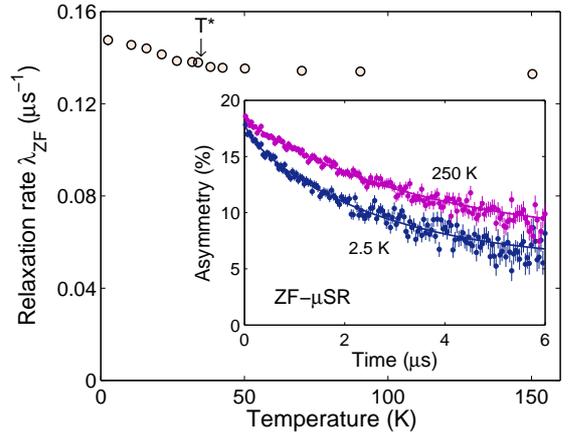} %{ThFeAsN_ZF_decay_rate}
\caption{\textbf{ZF-$\mu$SR time spectra and relaxation rates.} 
The zero-field relaxation rate $\lambda_{\mathrm{ZF}}$ is small and 
practically constant with temperature, with only a tiny increase below $T^*$. 
Inset: representative ZF-$\mu$SR spectra at selected temperatures.
}
\label{fig:ZF-MuSR}
\end{figure}

By means of systematic zero-field (ZF)-$\mu$SR measurements, sensitive 
to the material's intrinsic magnetic properties, we could follow the evolution 
of a possible magnetically-ordered phase. As shown in the inset of Fig.~\ref{fig:ZF-MuSR}, 
the time-domain $\mu$SR spectra generally exhibit exponential decays, which become 
more prominent as the temperature decreases. 
An ideal non-magnetic sample is expected to show a constant asymmetry. Real samples, 
however, invariably show small, mostly temperature-independent decays, attributed to 
nuclear magnetic moments or to tiny amounts of diluted ferromagnetic impurities, the 
latter commonly occurring in various IBS families \cite{Sanna2009JSNM,Lamura2014}. 
Their randomly oriented moments are known to create weak stray fields over the entire 
sample \cite{Walstedt1974}, hence giving rise to muon-spin relaxation.
In our case, magnetometry results are consistent with a 0.5\% extra amount of Fe 
(i.e., below the detection threshold of powder XRD) in the form of tiny dispersed clusters, 
which provide a temperature-independent average magnetic moment of 
ca.\ 0.03\,$\mu_{\mathrm{B}}$/Fe.
Accordingly, our ZF asymmetry data were analyzed by considering the sum of two contributions: 
$A_\mathrm{ZF} \-=\- A_\mathrm{ZF}^\mathrm{sample}\- + \-A_\mathrm{ZF}^\mathrm{imp.}$. 
The latter is relevant only at short times, but since its amplitude does not exceed 10\% of 
the total signal, it is hardly discernible in the inset of  Fig.~\ref{fig:ZF-MuSR}.  
The ZF-$\mu$SR signal of the sample is well described by exponential relaxations 
of the form (see inset in Fig.~\ref{fig:ZF-MuSR}):
\begin{equation}
\label{eq:ZF_fit}
A_\mathrm{ZF}(t)/A_\mathrm{ZF}(0) =  p_{\mathrm{fast}} e^{-\Lambda_{\mathrm{fast}} t} + p_{\mathrm{slow}} e^{-\Lambda_{\mathrm{slow}} t},
\end{equation}
where $p_{\mathrm{fast},\mathrm{slow}}$ and $\Lambda_{\mathrm{fast},\mathrm{slow}}$ 
are the relative weights and relaxation rates of muons implanted in two inequivalent sites, 
namely close to FeAs and to ThN planes, respectively \cite{Khasanov2008,Sanna2009}.
In agreement with these studies, we find $p_{\mathrm{slow}} \sim 15$\% at 
$T = 5$\,K, decreasing with temperature.

The relaxation rate $\lambda_\mathrm{ZF}$ ($\equiv \Lambda_{\mathrm{fast}})$, reflecting 
a possible magnetic order due to Fe$^{2+}$ ions, is shown in Fig.~\ref{fig:ZF-MuSR}.
It is practically independent of temperature, exhibiting a marginal increase 
(of only $\sim 0.02$\,$\mu$s$^{-1}$) below $T^* = 35$\,K (see NMR results in 
Fig.~\ref{fig:invT1T_shift} for a definition of $T^*$). Given the absence of 
applied magnetic fields in ZF-$\mu$SR experiments, an increase in 
$\lambda_\mathrm{ZF}$ is usually attributed to the onset of antiferromagnetic order 
[local moment or spin-density wave (SDW)].
However, given its tiny value and the absence of coherent muon-spin 
precession below $T^{*}$, it indicates a broad distribution of weak 
internal fields, i.e., no well-developed magnetic order in ThFeAsN. 
This is consistent with results of transport, magnetic \cite{Wang2016}, and 
${}^{57}$Fe M{\"o}ssbauer spectroscopy \cite{Albedah2017}
studies, where no magnetic order was detected down to 2\,K.
Similarly to the superconducting F- or H-doped 
 LaFeAsO \cite{Luetkens2009,Lamura2014}, 
as well as to many other 1111 compounds \cite{Sanna2010,Shiroka2011}, 
where a short-range magnetic order, vanishing with doping, is claimed 
to coexist with superconductivity, also in ThFeAsN the weak magnetism 
seems closely related to the onset of superconductivity.
Longitudinal-field $\mu$SR experiments (not discussed here) indicate 
that in ThFeAsN, too, the {weak magnetic moments behave as static} within 
the $\mu$SR time scale.
On the other hand, ThFeAsN clearly differs from pure LaFeAsO which, 
below $T_{\mathrm{N}}$, exhibits oscillating ZF-$\mu$SR signals \cite{Luetkens2009}, 
a signature of long-range magnetic order. From this comparison it is evident 
that the undoped ThFeAsN already fulfills the conditions to sustain 
superconductivity, which other 1111 compounds achieve only upon doping. 

%\vspace{2mm}
\subsection{Probing superconductivity via TF-$\mu$SR.} 
To explore the nature of superconductivity in ThFeAsN, we performed a series of 
transverse-field (TF)-$\mu$SR measurements from 1.6 to 35\,K.
A type-II superconductor exposed to an external magnetic field develops 
a regular flux-line lattice (FLL), which modulates the field distribution 
inside the material. Muons, which sample the FLL uniformly, experience an 
additional Gaussian relaxation $\sigma_{\mathrm{sc}}$, the latter being 
a measure of the absolute magnetic penetration depth $\lambda$ and, hence, 
related to the superfluid density $\rho_{\mathrm{sc}} \propto \lambda^{-2}$ 
\cite{Brandt2003,Brandt2009}.

\begin{figure*}[t]
\centering
\includegraphics[width=0.33\textwidth]{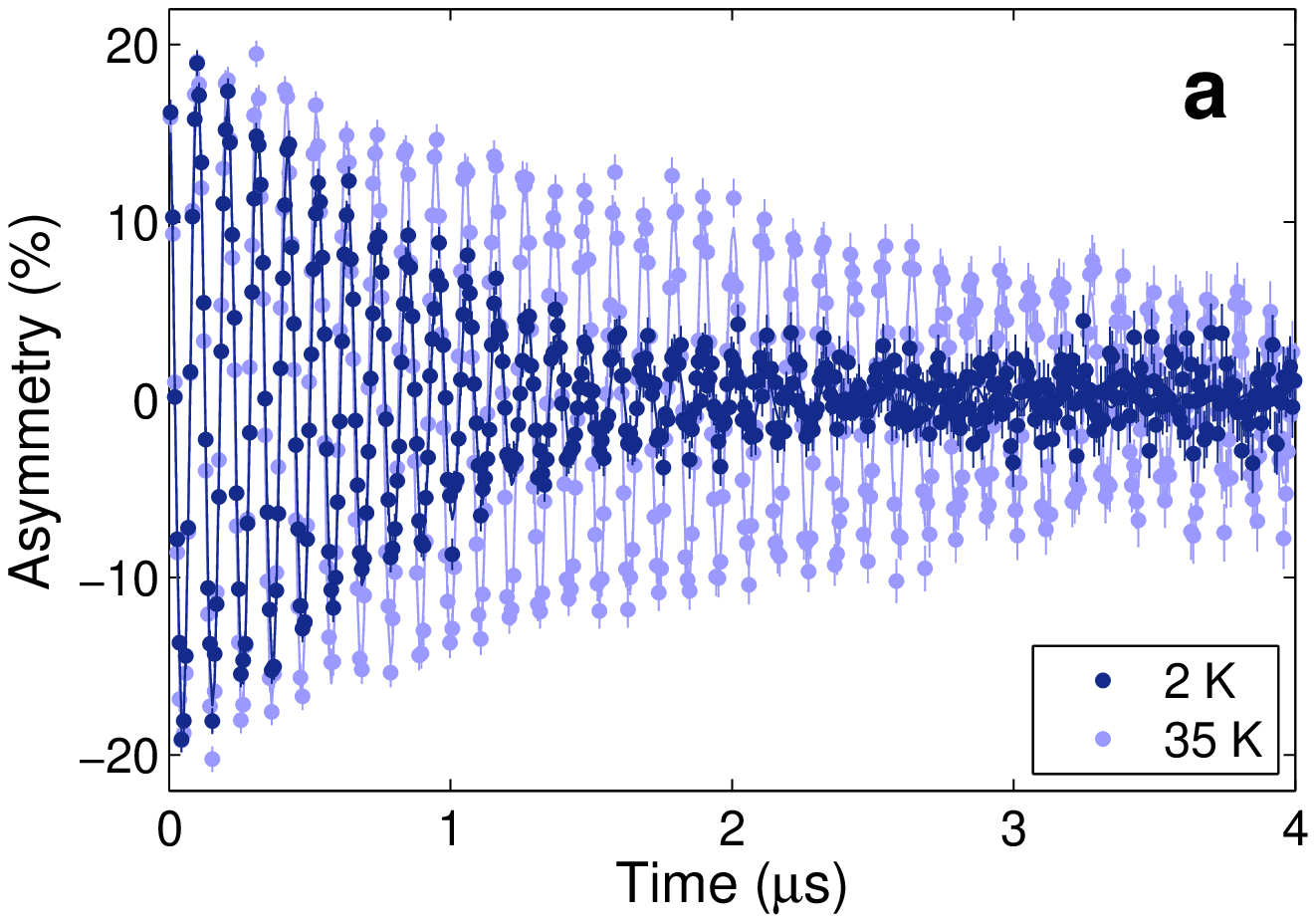} %{ThFeAsN_SC_TF_70mT_35_2K}
\includegraphics[width=0.325\textwidth]{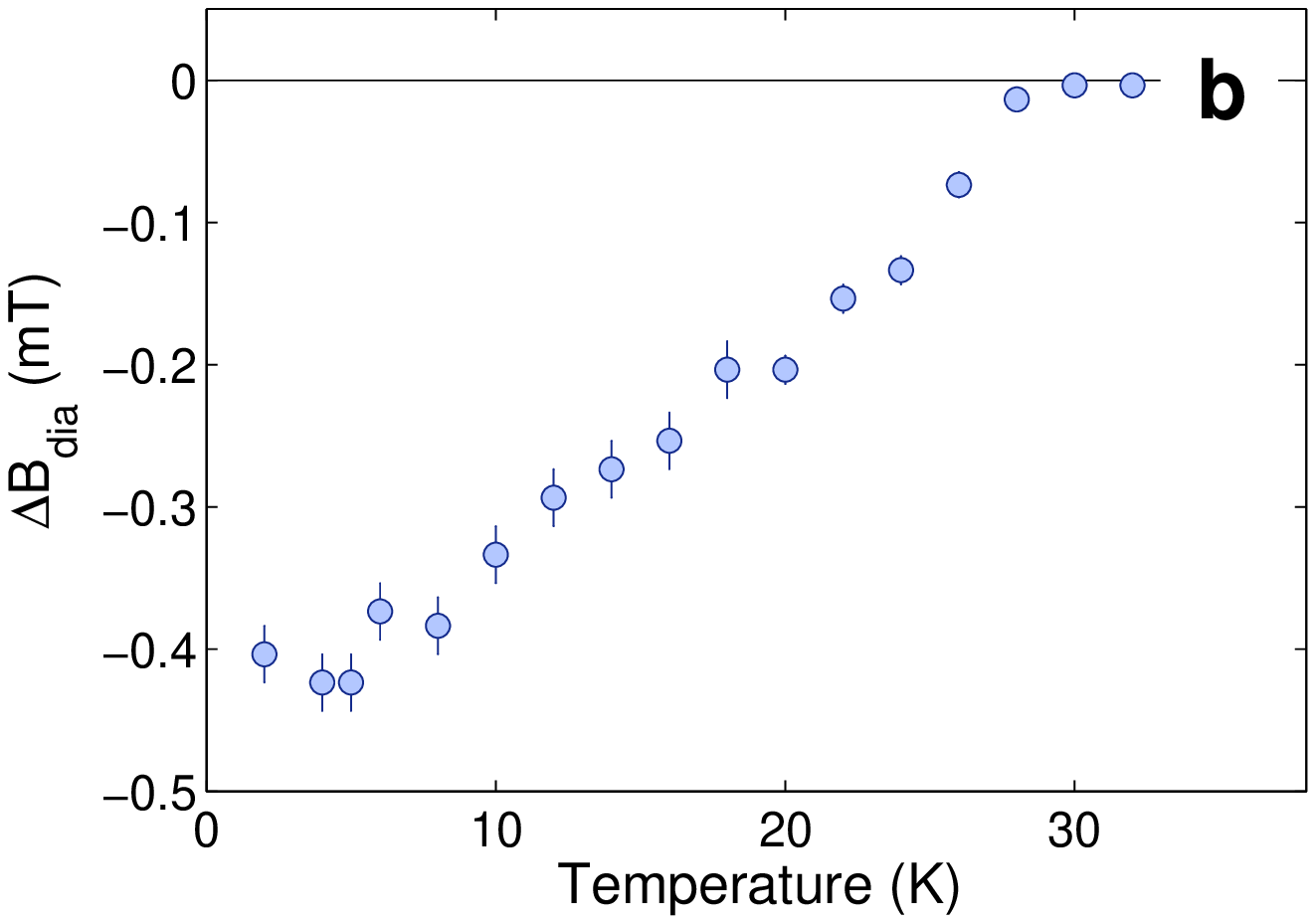} %{ThFeAsN_diamagn_TF_70mT}
\includegraphics[width=0.33\textwidth]{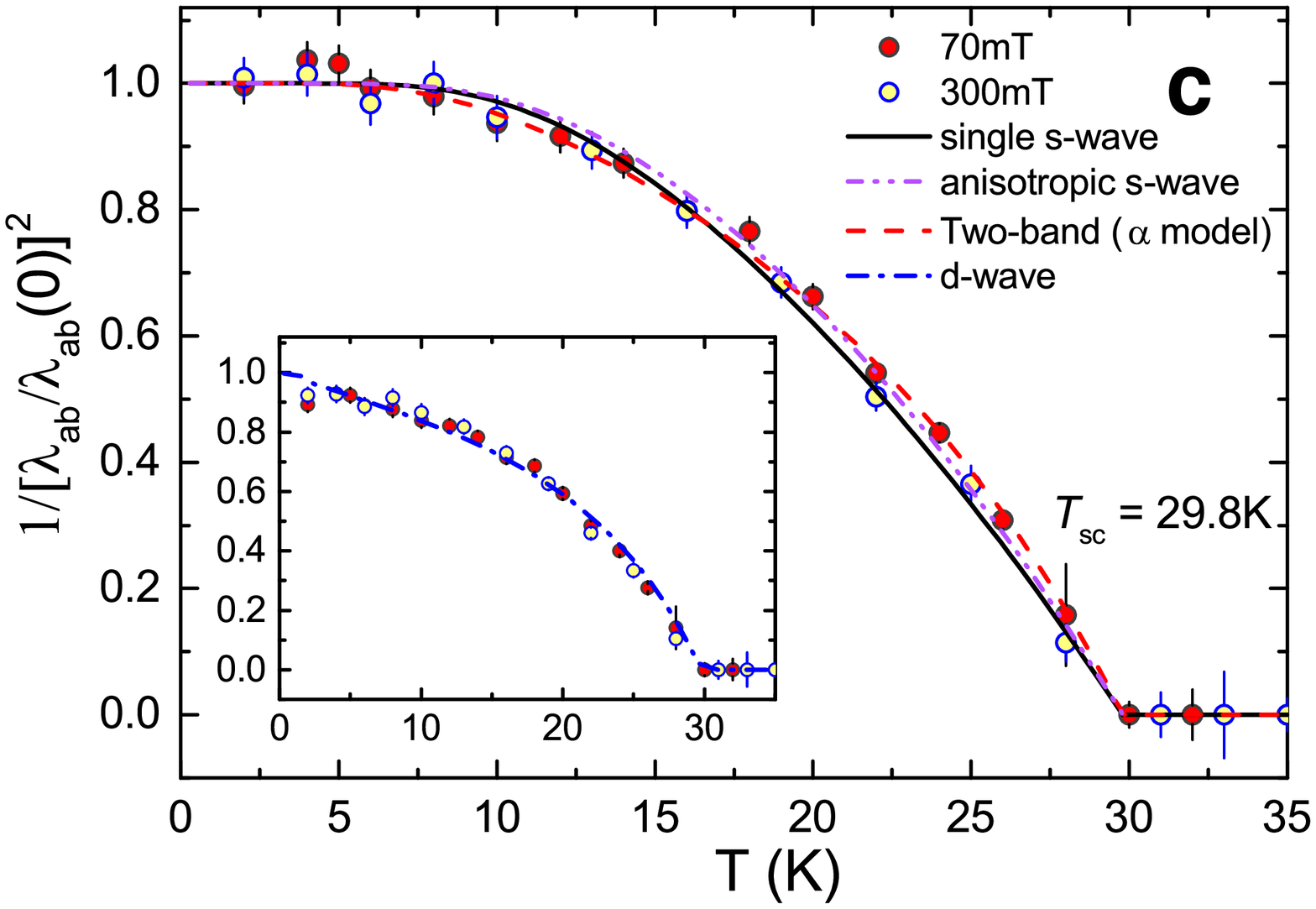} %{s-wave_gap_information}
\caption{\textbf{TF-$\mu$SR spectra, diamagnetic shift, and relaxation rate.}
(\textbf{a}) Representative TF-$\mu$SR spectra above and below $T_\mathrm{c}$
measured in 70\,mT and relevant fits.
(\textbf{b}) Diamagnetic field-shifts in the superconducting phase.
(\textbf{c}) Temperature dependence of $\lambda^{-2}$, as calculated from the measured 
TF relaxation rate $\sigma_{\mathrm{sc}}(T)$ at 70\,mT (red) and 300\,mT (yellow). 
Lines represent fits using a single-gap (solid) and two-gap or anisotropic $s$-wave 
model (dashed), the latter two showing a better $\chi^2_r$. The inset 
shows a fit using a $d$-wave model.
}
\label{fig:TF-MuSR}
\end{figure*}

The temperature dependence of $\sigma_{\mathrm{sc}}$ was studied in a 
70-mT transverse field under field-cooling conditions, with additional 
data collected also at 300\,mT. 
A preliminary field-dependence study of the $\mu$SR depolarization rate at 2\,K 
indicated that both fields are suitable for probing the intrinsic superconducting properties 
of ThFeAsN, since neither the decrease of the intervortex distance with field, nor the 
vortex-core effects \cite{Sonier2007} are of significance below $\sim 500$\,mT, 
considerably smaller than $B_{c2} \sim 50$\,T \cite{Cao2016}. 
Both datasets give comparable $\sigma_{\mathrm{sc}}$ values [see 
Fig.~\ref{fig:TF-MuSR}(c)] but, for a direct comparison with available data on 
other IBS compounds, we focus on the 70-mT case.
Figure~\ref{fig:TF-MuSR}(a) shows two typical TF-$\mu$SR spectra measured
above and below $T_\mathrm{c}$, fitted by means of:
\begin{equation}
A_\mathrm{TF} = A_\mathrm{TF}(0) \, \cos(\gamma_{\mu} B_{\mu} t + \phi) e^{-\lambda_\mathrm{ZF} t} e^{- \sigma^2 t^2/2}.
\end{equation}
Here $A_\mathrm{TF}(0)$ is the initial asymmetry, $\gamma_{\mu} = 2\pi \times 135.53$\,MHz/T
is the muon gyromagnetic ratio, $B_{\mu}$ is the local field sensed by the implanted muons, $\phi$ 
is the initial phase, $\lambda_\mathrm{ZF}$ is a Lorentzian-, and $\sigma$ a Gaussian-relaxation 
rate. As follows from the above ZF analyses [see Fig.~\ref{fig:ZF-MuSR}], the Lorentzian 
relaxation reflects coexisting magnetic correlations and is significantly smaller than the 
SC-dominated Gaussian relaxation rate, $\sigma$. The latter comprises contributions from both 
the FLL ($\sigma_\mathrm{sc}$) and a small temperature-independent relaxation due to nuclear moments ($\sigma_\mathrm{n}$), 
determined above $T_\mathrm{c}$. Below $T_\mathrm{c}$, the FLL-related relaxation was obtained by subtracting 
the nuclear contribution from the Gaussian relaxation rate, i.e., 
$\sigma_\mathrm{sc}^2 = \sigma^2 - \sigma_\mathrm{n}^2$. 

Figure~\ref{fig:TF-MuSR}(b) depicts the diamagnetic shift below $T_\mathrm{c}$, whose value 
increases deep inside the superconducting phase. The simultaneous development of a 
flux-line lattice at low temperatures implies the appearance of $\sigma_\mathrm{sc}$, 
in turn reflecting an increase in $1/\lambda^2$ [see Fig.~\ref{fig:TF-MuSR}(c)]. 
For small applied fields [in comparison with $B_{c2}(0)$] and hexagonal 
flux-line lattices the two quantities are related by 
\cite{Barford1988,Brandt2009}:
\begin{equation}
\frac{\sigma_\mathrm{sc}^2(T)}{\gamma^2_{\mu}} = 0.003\,71\, \frac{\phi_0^2}{\lambda^4_{\mathrm{eff}}(T)},
\end{equation}
with $\lambda_{\mathrm{eff}}$ the effective penetration depth. 
We recall that in anisotropic polycrystalline superconducting samples (as is the case
of the layered ThFeAsN compound) the effective penetration depth $\lambda_{\mathrm{eff}}$
is determined mostly by the shortest penetration depth $\lambda_\mathrm{ab}$, the exact relation 
between the two being $\lambda_{\mathrm{eff}} = 3^{1/4}\lambda_\mathrm{ab}$ \cite{Fesenko1991}.

Figure~\ref{fig:TF-MuSR}(c) shows the temperature dependence of the normalized 
superfluid density, $\rho_{\mathrm{sc}} \propto \lambda_\mathrm{ab}^{-2}$.
The temperature-independent behavior of $\lambda_\mathrm{ab}^{-2}$
below 7\,K indicates a fully gapped superconductor, thus excluding, e.g., a 
$d$-wave gap structure. Indeed, as shown in the inset of Fig.~\ref{fig:TF-MuSR}(c), 
a $d$-wave model does not properly fit the experimental data, especially below 5\,K. 
However, as can be seen in the main panel of Fig.~\ref{fig:TF-MuSR}(c), 
single $s$-wave, two-gap $s$-wave or anisotropic $s$-wave 
gap models are all compatible with the observed $\lambda_\mathrm{ab}^{-2}(T)$ behavior.
The better agreement of the two-gap model (or $s-$wave gap with substantial 
anisotropy) with the data is confirmed by the slightly positive 
curvature in $B_{c2}(T)$ close to $T_\mathrm{c}$, as derived from magnetization and 
electrical resistivity measurements, which also favor a multigap SC state. 
Given the two to five Fe-related bands crossing the Fermi surface, the occurrence of 
multiple SC gaps in iron-based superconductors is not surprising \cite{Si2016,Chubukov2008}.
Recent Fermi surface calculations \cite{WangG2016,Singh2016} suggest a similar 
scenario also for ThFeAsN.

A comparison of different families of superconductors can be summarized in a so-called 
Uemura plot \cite{Uemura1991}. This type of representation tracks the dependence 
of $T_\mathrm{c}$ on the inverse square of the in-plane London penetration depth, $\lambda_\mathrm{ab}^{-2}$.
Since $1/\lambda_\mathrm{ab}^2(0) \sim \rho_{\mathrm{sc}}/m^{\star}$, with
$\rho_{\mathrm{sc}}$ the superfluid density and $m^{\star}$ the renormalized
mass of the quasiparticles, a positive correlation between $T_\mathrm{c}$ and $\rho_{\mathrm{sc}}/m^{\star}$ 
was identified. In this diagram, ThFeAsN lies close to the dataset 
for FeSe and, remarkably, also to the electron-doped LaFeAsO.
This is not surprising given the almost identical 
results of the first-principle electronic-structure calculations for ThFeAsN 
and LaFeAsO \cite{WangG2016,Singh2016}.
In view of this, the observation of superconductivity in stoichiometric 
ThFeAsN without any additional doping is quite remarkable.

\subsection{Complementary findings from NMR investigations.}
As a complementary technique to $\mu$SR, ${}^{75}$As NMR was used to 
locally probe the  static (line widths and shifts) and the dynamic (spin-lattice relaxation) 
properties of ThFeAsN. The results of these measurements (at 7\,T) 
confirm the features presented and discussed above, but provide 
also additional insight into the characteristics of superconductivity of ThFeAsN.

\begin{figure}[t]
\includegraphics[width=0.9\columnwidth]{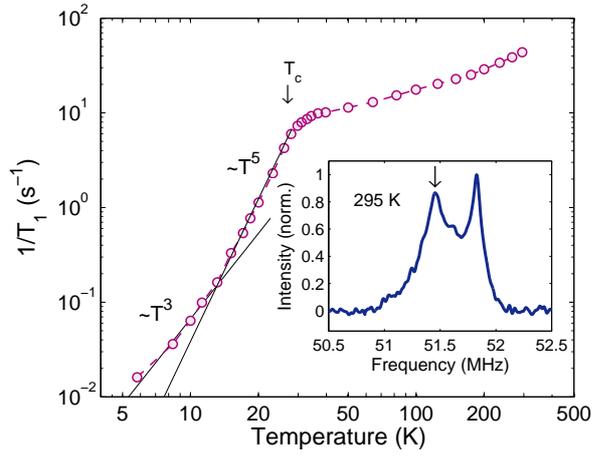} %{ThFeAsN_invT1_lineshape}
\caption{\label{fig:relaxations}\textbf{NMR relaxation rate and lineshape.}
Temperature dependence of $1/T_1$ relaxation rate 
measured at 7\,T at the left peak of the ${}^{75}$As NMR lineshape (arrow in the inset). 
A similar $1/T_1$ behavior is found also for the right peak. The steep decrease of $1/T_1$ 
below $T_\mathrm{c}$ is compatible with a fully-gapped superconductor. Lines indicate different 
power-law dependences, at different temperature regimes (see text for details). 
}
\end{figure}

Since ${}^{75}$As has a nuclear spin $I = 3/2$, the observed NMR line is broadened 
by a second-order quadrupole perturbation of the central Zeeman $+1/2$ to $-1/2$ 
transition, the satellites being much weaker and far apart. Both the two-peak lineshape 
(see inset in Fig.~\ref{fig:relaxations}) and its variation with temperature 
are very similar to those of the ${}^{75}$As NMR lines observed in lightly F-doped 
LaFeAsO \cite{Grafe2008,Nakai2009}. Most importantly, consistent with 
the $\mu$SR results, we observe only a marginal increase (less than 5\%) in 
FWHM below 35\,K, hence confirming the absence of magnetic order.
The resonance frequencies can be modeled by $f = \gamma H_0 (1 + K) + f_Q$, 
with $K$ the magnetic shift due to interactions between the probe nucleus and 
its electronic environment, and $f_Q = -771$\,kHz a temperature-independent 
second-order quadrupole shift. 

\begin{figure}[thb]
\includegraphics[width=0.9\columnwidth]{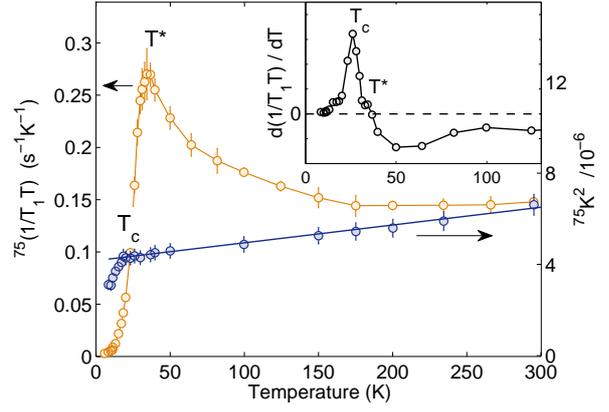} %{ThFeAsN_invT1T_shift}
\caption{\textbf{Different temperature behavior of $1/(T_1T)$ and $K^2$.}
Temperature dependence of $1/(T_1T)$ (left scale) and ${}^{75}K^2$ 
shift (right scale) measured at 7 T. The peak in the former and the drop in the 
latter indicate $T^{*}$ and $T_\mathrm{c}$, respectively, which differ by ca.\ 5\,K.
The clearly different functional form of the two curves below ca.\ 150\,K indicates 
the development of strong AF fluctuations. The strong low-temperature 
drop of $1/(T_1T)$ confirms the bulk character of superconductivity, 
whereas the peak in the derivative (inset) indicates its sharp onset.
}
\label{fig:invT1T_shift}
\end{figure}

Our most intriguing result is captured in Fig.~\ref{fig:invT1T_shift}, 
which compares the temperature dependencies of $(T_1T)^{-1}$ and the squared 
magnetic shift $K^2$. Although according to standard theory, $(T_1T)^{-1} \propto K^2$ 
\cite{Slichter1992}, this is obviously not the case here. In the normal state, $K^2$ 
decreases weakly and linearly with decreasing temperature. $(T_1T)^{-1}(T)$, 
however, starts growing steadily below approximately 175\,K 
[because of the chosen logarithmic scales in Fig.~\ref{fig:relaxations}, this growth 
is less distinct, yet clearly recognizable also in $T_1^{-1}(T)$ data.].
This relative increase in relaxation is abruptly terminated at $T^* = 35$\,K, followed 
by a steep decrease, masking the onset of superconductivity at 
$T_\mathrm{c}(7\,\mathrm{T}) \approx 27$\,K (identified by independent resistivity measurements). 
Note that a similar behavior of $(T_1T)^{-1}$ was observed also in the 
lightly-doped LaFaAsO$_{1-x}$F$_x$ \cite{Nakai2009} 
or in the isoelectronically-substituted LaFaAs$_{1-x}$P$_x$O class \cite{Shiroka2017}. 
As argued in these cases, we also tentatively ascribe the extra relaxation 
to the growing influence of spin fluctuations. However, in the above 
examples, spin-fluctuations precede an incoming AF order, in turn a precursor of superconductivity.
By contrast, in ThFeAsN (where no AF order is detected) we postulate 
the opening of a gap in the corresponding spin-excitation spectrum 
at $T^*$, exceeding $T_\mathrm{c}$, the latter clearly indicated by the sudden drop 
in $K^2(T)$ at the onset of superconductivity.
These data suggest the onset of a superconducting state in competition 
with spin fluctuations.

Once the superconducting state is established below $T_\mathrm{c} $, a gradual 
decrease of $K$ (see Fig.~\ref{fig:invT1T_shift}) hints at a spin-singlet pairing, for 
which the coupling between opposite spins implies a reduction of the local spin susceptibility.
The temperature dependence of $T_1^{-1}$ (see Fig.~\ref{fig:relaxations}) reflects 
a combination of growing gaps in the spectra of both spin- and electronic excitations. 
Interesting is the transient $T^5$-regime for $T_1^{-1}(T)$, just below $T_\mathrm{c}$, 
where the spin gap is not yet fully developed. 

\section{Discussion}
One of the most intriguing questions regarding ThFeAsN is the occurrence 
of bulk superconductivity in a stoichiometric compensated-metal compound 
without long-range magnetic order.
Because of a strong nesting of the electron- and hole bands (separated by the AF momentum), 
band-structure theory predicts that both undoped ThFeAsN and LaFeAsO should show a similar 
($C$-type) stripe antiferromagnetic order \cite{WangG2016,Singh2016}.
LaFeAsO indeed adopts a magnetically ordered ground state, 
while, as we have shown, ThFeAsN does not, but exhibits rather strong magnetic fluctuations. 
This key feature indicates a sizable renormalization of the electronic structure of ThFeAsN, 
as compared to that of LaFeAsO, beyond the density-functional theory. 
The simultaneous presence of fluctuations and the lack of magnetic order puts the stoichiometric 
compensated metal ThFeAsN into the same category as LiFeAs or FeSe, which also show strong 
correlations, although these correlations do not give rise to a magnetically-ordered ground state, 
as instead is the case of LaFeAsO, CaFe$_2$As$_2$, etc. 
We recall also that in LiFeAs and FeSe the absence of magnetic order 
allows for an orbital-selective superconducting state to be realized, 
which involves a strong orbital renormalization and differentiation with respect 
to Cooper-pairing on different orbitals. This contrasts with a conventional $s^{\pm}$ 
superconducting state, currently assumed for the F-doped LaFeAsO and 
K- or Co-doped BaFe$_2$As$_2$, where the gap is less orbital-dependent 
and simply changes sign between all-electron and all-hole pockets.

Clearly, the undoped ThFeAsN shares with optimally F-doped LaFeAsO both 
the presence of magnetic fluctuations and the occurrence of superconductivity. 
Yet, their similarity is only apparent. Indeed, while in F-doped LaFeAsO the 
suppression of the original magnetic order is simply caused by doping, the mechanism 
of magnetic-order suppression in stoichiometric 
ThFeAsN has another, not yet known, origin. The missing information 
on the electronic renormalization in ThFeAsN can be provided by future 
ARPES or FT-STM measurements, once single crystals would be available. 
In any case, ThFeAsN currently offers the unique opportunity of studying 
the peculiarities of unconventional superconductivity in the correlated 
iron-based superconductors.

\vspace{4mm}
In conclusion, our results establish the proximity of superconductivity 
with a competing state exhibiting sizable spin fluctuations in ThFeAsN. 
For many other iron-pnictide superconductors it has been shown that 
magnetic order and superconductivity coexist microscopically 
(see, e.g., \cite{Sanna2010,Shiroka2011}). 
As outlined above, this is certainly not the case for ThFeAsN, 
whose lack of magnetic order places it in the same class with LaFePO, LiFeAs, and FeSe.
Yet, as shown by NMR relaxation data, the proximity of a gapped spin-fluctuating 
phase to SC, does not exclude a spin-fluctuation meditated 
SC pairing in ThFeAsN.
Microscopic $\mu$SR and NMR measurements indicate also that 
the electronic excitation spectrum in the superconducting state 
is best described by a two-gap $s$-wave (or an anisotropic $s$-wave) model.

In a broader context, it was established that, e.g., in BaFe$_2$As$_2$, key 
structural features, such as the Fe-Fe distance and the As-Fe-As bond angle, vary in the same 
way under pressure or upon chemical doping, inducing similar electronic-structure 
evolutions in both cases \cite{Kimber2009}.
While for most 1111 and 122 families the required electronic-structure 
modifications to sustain superconductivity are achieved via chemical-doping, 
for ThFeAsN an appropriate combination of structural and electronic parameters 
results in an enhanced $T_\mathrm{c}$ already in its undoped state.
This is not surprising, considering its $c/a$ ratio of 2.11 \cite{Wang2016}, 
significantly reduced in comparison with other 1111 and 122 compounds, 
a difference ultimately due to the different ionic sizes of N${}^{3-}$ (1.46\,\AA) 
and O${}^{2-}$ (1.38\,\AA).
The ThFeAsN case suggests that  Fermi-surface modifications by structural 
distortions and correlation effects are as important as charge doping in inducing 
superconductivity in IBS compounds. 
Yet, the structural route to SC is so rare, because very few compounds 
do exhibit a suitable combination of structural parametes.
Indeed, our recent high-pressure measurements have shown a sizable 
decrease of $T_\mathrm{c}$ with increasing pressure \cite{Barbero2017}, 
confirming that ThFeAsN has already the optimal parameters to achieve the 
highest $T_\mathrm{c}$.

\vspace{5mm}

\section{Methods}
\subsection{Sample preparation.}
\footnotesize{%
Polycrystalline ThFeAsN specimens with typical grain sizes 
1--5\,$\mu$m were synthesized by solid-state 
reaction methods, as recently described in Ref.~\onlinecite{Wang2016}. 
The purity of precursors was checked via X-ray diffraction (XRD), while 
the composition of the final product was determined via EDX spectroscopy. 
The room-temperature XRD of ThFeAsN reveals a tetragonal 
structure ($P4/nmm$) with $a = 4.037$\,\AA\ and $c =  8.526$\,\AA\ (see inset in 
Fig.~\ref{fig:magnetization}) and no detectable impurity phases. 
The precise determination of the N content is rather challenging. While 
the possibility of N deficiencies cannot be absolutely ruled out, the fact 
that oxygen vacancies in our 1111 system are stabilized only under 
high-pressure synthesis suggests that N deficiencies, if present, 
are negligible.
}

\subsection{$\mu$SR and NMR experiments.} 
\footnotesize{%
Experiments employing $\mu$SR were done at the GPS spectrometer of the Paul 
Scherrer Institute (PSI) in Villigen, Switzerland. Once implanted in matter, spin-polarized 
muons ($\mu^+$) act as microscopic probes of the local magnetic environment, which 
upon decay emit positrons preferentially along the muon-spin direction. The spatial 
anisotropy of the emitted positrons (i.e., the asymmetry signal) reveals the 
distribution of the local magnetic fields at the muon site \cite{Blundell1999,Yaouanc2011}.
As for the NMR investigations, a broad-band spectrometer was used for the static 
(line-widths and -shifts) and the dynamic (spin-lattice relaxation) measurements. 
With a 100\% isotopic abundance and a relatively large gyromagnetic ratio, 
the ${}^{75}$As nucleus ($I = 3/2$) was the probe of choice.
In case of $\mu$SR measurements, the error bars in the raw data were obtained 
from the counting statistics, while for the NMR from the noise levels and the frequency 
resolution. All the other error bars were calculated by using the standard methods 
of error propagation. 
}

\subsection{Extracting superconducting parameters from $\mu$SR data.} 
\footnotesize{%
The two-gap $s$-wave model provides $\lambda_\mathrm{ab}(0) = 230(2)$\,nm, with the
two superconducting gap values being $\Delta_1(0) = 3.4(2)$\,meV and 
$\Delta_2(0) = 6.5(3)$\,meV, with weighting factors $w_1 = 0.40(2)$ and 
$w_2 = 0.60(3)$, respectively. From these we find the following gap-to-$T_\mathrm{c}$ ratios: 
$\Delta_1(0)/k_\mathrm{B} T_\mathrm{c} = 1.3$ and $\Delta_2(0)/k_\mathrm{B} T_\mathrm{c} = 2.5$,
with $k_{\mathrm{B}} = 8.62 \times 10^{-2}$\, meV/K the Boltzmann constant 
and $T_\mathrm{c} = 29.8$\,K. Similar values are obtained in the anisotropic 
single-gap $s$-wave case: $\lambda_\mathrm{ab}(0) = 250(4)$\,nm, 
$\Delta(0) = 4.8(5)$\,meV, and $\Delta(0)/k_\mathrm{B} T_\mathrm{c} = 1.9$.
Since for an ideal BCS superconductor the last ratio is expected to be 1.76, we 
conclude that ThFeAsN is a superconductor in the weak-coupling limit.
The above values are in principle close to those of F-doped LaFeAsO, with $T_\mathrm{c} = 24$\,K 
and $\Delta(0) = 3.6$\,meV \cite{Sato2008}.

From the results of magnetometry and $\mu$SR measurements, other relevant SC 
parameters for ThFeAsN can be extracted. By using the formula $B_{c2} = \phi_0/(2\pi\xi^2)$, 
where $\phi_0 = 2.07\times10^{-15}$\,Wb is the magnetic flux quantum, we 
estimate a superconducting coherence length $\xi(0) = 2.57$\,nm. This rather 
small $\xi$ value, combined with a large penetration depth $\lambda_\mathrm{ab}(0) = 230(2)$\,nm, 
as determined via TF-$\mu$SR, indicate that ThFeAsN is an extreme type-II superconductor, 
with a Ginzburg-Landau parameter $\kappa = \lambda/\xi \simeq 90$.
By using this value for $\kappa$ and the formula 
$B_{c1}(0) = \phi_0/(4\pi \lambda_\mathrm{ab}^2)\ln{\kappa}$ \cite{Buckel2004}, 
we also obtain the lower critical field $B_{c1}(0) = 13.4$\,mT, similar in magnitude 
to the $B_{c1}$(0) values of other IBS compounds \cite{Johnston2010}.
}

\subsection{Data availability.} 
\footnotesize{%
The data that support the findings of this study are available 
from the corresponding authors upon reasonable request.
}

\section{Acknowledgments} 
\footnotesize{%
Part of this work was performed at the Swiss Muon Source (S$\mu$S) Paul 
Scherrer Institut, Villigen, Switzerland. E.\ Morenzoni and A.\ Amato are acknowledged 
for providing fast-track access to the GPS instrument.
The authors thank S.\ Holenstein and R.\ Khasanov for the assistance during 
the experiments and F.\ Lochner and J.\ L.\ Zhang for useful discussions.
This work was financially supported in part by the Schweizerische Nationalfonds
zur F\"{o}rderung der Wissenschaftlichen Forschung (SNF) and the National 
Natural Science Foundation of China (Grant No. 11304183).
I.E.\ acknowledges support from the project of the state assignment of KFU in the 
sphere of scientific activities (Grant No.\ 3.2166.2017/4.6).
}

\section{Author contributions}
\footnotesize{%
Project planning: T.\ Shiroka. Sample growth: C.W.\ and G.H.C. 
$\mu$SR experiments were carried out by T.\ Shiroka and T.\ Shang. 
T.\ Shang performed also the $\mu$SR data analysis. NMR measurements 
and analysis: T.\ Shiroka. Theoretical input: I.E. 
The manuscript was drafted by T.\ Shiroka and H.R.O.\ and was completed 
with input from all the authors.
}

\footnotesize{%
\vspace{2mm}
\noindent\textbf{Competing financial interests:} The authors declare 
no competing financial interests.
}

\vspace{-5mm}
%After running bibtex, copy the contents of the .bbl file into the original document.
%\begin{thebibliography}{99}
\renewcommand*{\bibfont}{\footnotesize}
%\bibliography{biblio_Th}
%\end{thebibliography}


\end{document}